# Hot electron dynamics in a semiconductor nanowire under intense THz excitation

Andrei Luferau, Maximilian Obst, Stephan Winnerl, Alexej Pashkin, Susanne C. Kehr, Emmanouil Dimakis, Felix G. Kaps, Osama Hatem, Kalliopi Mavridou, Lukas M. Eng, and Manfred Helm

**Abstract:** We report THz-pump / mid-infrared probe near-field studies on Si-doped GaAs-InGaAs core-shell nanowires utilizing THz radiation from the free-electron laser FELBE. Upon THz excitation of free carriers, we observe a red shift of the plasma resonance in both amplitude and phase spectra, which we attribute to the heating up of electrons in the conduction band. The simulation of heated electron distributions anticipates a significant electron population in both L- and X-valleys. The two-temperature model is utilized for a quantitative analysis of the dynamics of the electron gas temperature under THz pumping at various power levels.

High-quality epitaxial nanowires (NWs) based on III–V semiconductors offer the possibility to fabricate ultrafast optical devices. In particular, single NWs can serve as sensitive element of terahertz (THz) radiation detectors that operate at room temperature [1, 2]. Contactless investigation of the average charge carrier concentration and mobility in large ensembles of NWs is possible using THz time-domain spectroscopy [3]. The local investigation of these properties on individual NWs can be carried out by scattering-type scanning near-field optical microscopy (s-SNOM) [4]. This technique offers a spatial resolution far beyond the diffraction limit, and in case of far-infrared studies, it can be extended beyond $\lambda/4600$ [5, 6]. Utilizing near-infrared (NIR) excitation on (typically *intrinsic*) NWs, charge-carrier lifetimes in NW ensembles become accessible in far-field NIR-pump / THz-probe experiments [7, 8]. Several studies demonstrated the feasibility of integrating NIR-pump / THz-probe technique with near-field microscopy for the examination of III–V semiconductors [9-12]. NIR pump served as *interband* excitation and enabled to investigate the local recombination dynamics of photogenerated carriers in individual NWs. However, for some important applications of NWs such as field-effect transistors [13-15] or THz detectors [1, 2] it is important to study the response of *doped* NWs to intense THz radiation that does not change the carrier density, but rather heats up the carrier system via *intraband* absorption.

Here we report on THz-pump / mid-infrared (MIR) probe s-SNOM studies on highly-doped GaAs/ InGaAs core-shell NWs utilizing the intense narrowband THz radiation from the free-electron laser (FEL) FELBE at the Helmholtz-Zentrum Dresden-Rossendorf [16]. The analysis of the near-field spectra enables us to obtain the electron temperature as a function of the pump-probe delay time. Our results quantify the dynamics of hot electrons in the conduction band of the InGaAs shell and enable us to quantify the electron-phonon coupling. Furthermore, we get an insight into the electron heating and cooling dynamics by simulating it with the two-temperature model.

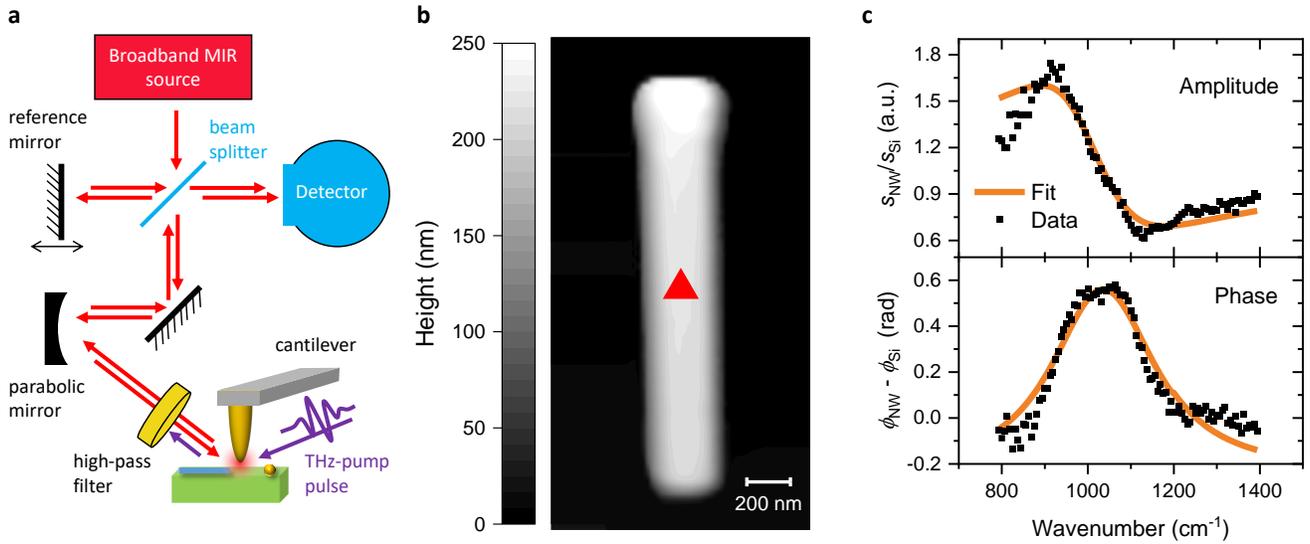

**Fig. 1.** (a) Sketch of time-resolved near-field spectroscopy setup based on scattering scanning near-field optical microscope (s-SNOM). A Michelson interferometer enables mid-infrared near-field spectroscopy (nanoFTIR) during terahertz pumping of the sample (b) Topography of the GaAs/InGaAs core-shell nanowire recorded by AFM. The tip position chosen for spectrally resolved scans is highlighted with a red triangle. (c) Near-field amplitude $s(\omega)$ and phase $\phi(\omega)$ spectra of doped GaAs/InGaAs core-shell nanowire normalized to the response of Si along with the results of two-parameter fit based on the point-dipole and the Drude models.

## RESULTS AND DISCUSSION

### Near-field MIR Spectroscopy of the Nanowires

The samples under study are Si-doped GaAs/InGaAs core-shell NWs grown by molecular beam epitaxy. They consist of a 25-nm-thick GaAs core and an 80-nm-thick $In_{0.44}Ga_{0.56}As$ shell that is homogeneously $n$-doped with a nominal Si-concentration of about $9 \times 10^{18}$ cm$^{-3}$ [17, 18]. For s-SNOM studies, these NWs are transferred onto a (100) Si substrate and dispersed randomly across it.

Our experiments are carried out with an s-SNOM setup from Neaspec GmbH equipped with a nanoFTIR module, including a broadband MIR difference-frequency generation (DFG) source (5 – 15 µm; 20 – 60 THz; $P_{avg}$~0.2 mW; 78 MHz; t <100 fs) used as a probe [19]. Radiation is focused by an off-axis parabolic mirror (numerical aperture: 0.46, focal length: 11 mm) onto a metallized atomic-force-microscope (AFM) tip that operates in tapping mode with an oscillation amplitude of about 100 nm. Tip-sample near-field interaction modifies the backscattered radiation, which is detected by a photoconductive mercury cadmium telluride (MCT) detector. To separate the near-field signal from the far-field, the detector signal is demodulated by a lock-in amplifier at the 2$^{nd}$ harmonic of the AFM cantilever tapping frequency $\Omega \approx 250$ kHz [20]. The tip and the sample are located in one of the arms of an asymmetric Michelson interferometer of the nanoFTIR module [Figure 1(a)]. Here, the tip-scattered near-field light optically interferes with the split MIR beam from the reference arm, and the detector signal is recorded as a function of the reference mirror position [21]. Finally, the obtained interferograms are Fourier transformed to get spectra of the near-field scattering amplitude $s(\omega)$ and phase $\phi(\omega)$.

In Figure 1(b), the AFM topography image of such a NW under study is shown, highlighting the chosen tip position for spectrally resolved scans with a red triangle. The orientation of NWs with respect to the incident angle of the laser shows no significant impact on the spectral response, while positioning the tip at the center of the NW prevents geometry-related artifacts like interference or shadowing.

A point-dipole model incorporating the frequency-dependent infrared permittivity $\varepsilon(\omega)$ of the NWs is used to estimate the near-field response of the samples [22, 23]. In this model the field $E_{scat}$ scattered by the tip is approximated by a dipolar near-field interaction between the tip apex and the sample surface resulting in:

$$E_{scat} \propto \frac{\alpha_0(1+r_p)^2}{1-\dfrac{\alpha_0\beta}{16\pi(a+h)^3}} E_{inc}, \tag{1}$$

where $\alpha_0 = 4\pi a^3$ is the polarizability of the tip with a radius $a$ induced by the incident field $E_{inc}$, $r_p$ is the Fresnel reflection coefficient for p-polarized light, $h$ is the tip apex–sample separation, and $\beta = (\varepsilon - 1)/(\varepsilon + 1)$. Consideration of higher harmonic demodulation involves incorporating a sinusoidal tip oscillation $h(t)$. The subsequent Fourier transformation of the $E_{scat}[h(t)]$ results in $E_n \propto s_n e^{i\phi_n}$, where $s_n$ and $\phi_n$ represent the amplitude and phase of the scattered electric field at demodulation order $n$.

The frequency dependence of the permittivity $\varepsilon(\omega)$ in a doped semiconductor can be described by the Drude model as following:

$$\varepsilon_{NW}(\omega) = \varepsilon_{optic} - \frac{\omega_{pl}^2 \varepsilon_{optic}}{\omega^2 + i\omega\gamma_{el}}, \tag{2}$$

where $\varepsilon_{optic}$ is the high-frequency permittivity, and the second term in the equation describes the response of free charge carriers with the plasma frequency $\omega_{pl}^2 = ne^2/(m^*\varepsilon_0\varepsilon_{optic})$ and the plasmonic damping $\gamma_{el} = e/(\mu m^*)$, where $e$ is the elementary charge, $n$ is the concentration of the charge carriers, $m^*$ and $\mu$ their effective mass and mobility, respectively. Here, we neglect the contribution from the polar phonon resonance since it is located well below the frequency range addressed in this study. Since the InGaAs shell is much thicker than the GaAs core, we neglect the core-shell structure and estimate the NW response as the plasmonic response of the heavily doped shell only.

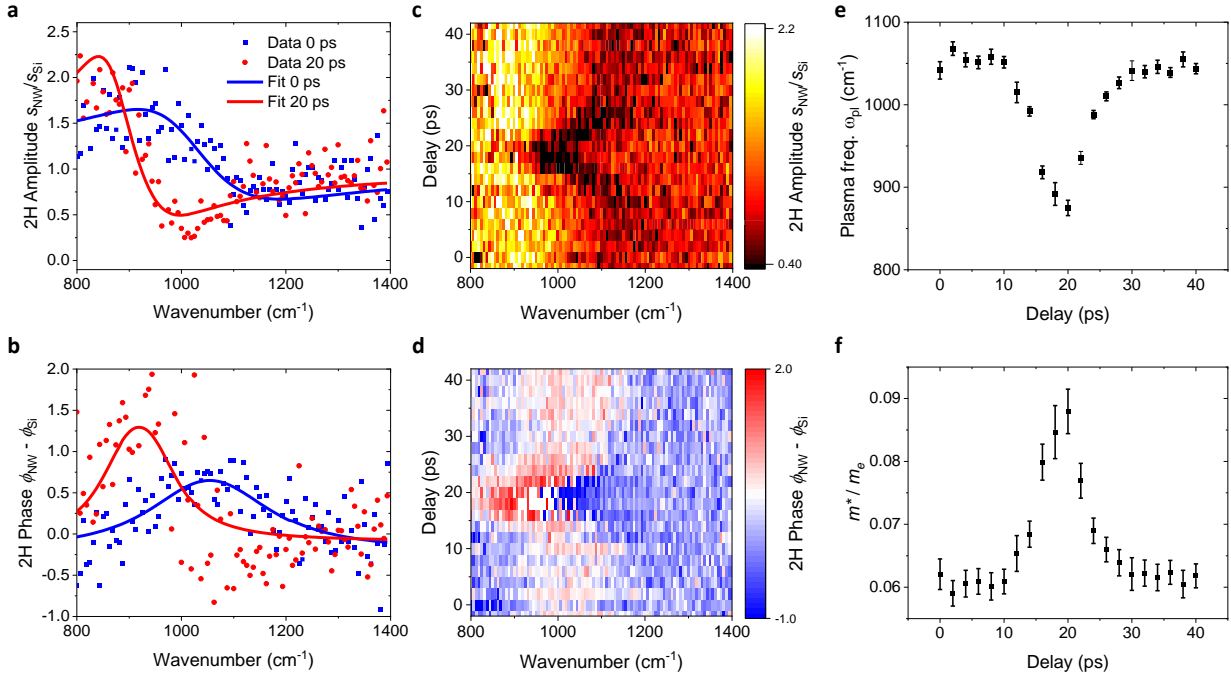

**Fig. 2.** (a, b) the Near-field amplitude s(ω) and phase ϕ(ω) spectra of doped GaAs/InGaAs core-shell nanowire (NW) obtained with (red) and without (blue) terahertz (THz) pumping (6 mW), normalized to the response of Si. The spectra are extracted from experimental data according equation (3) and plotted along with the results of two-parameter fit based on the point-dipole and the Drude models. (c,d) Color maps illustrating the evolution of the near-field amplitude s(ω) and the phase ϕ(ω) spectra of doped InGaAs NW upon THz photoexcitation (6 mW), normalized to the response of Si. Every line represents normalized near-field spectra obtained for different time delays between THz-pump and broadband mid-infrared probe. (e) Fitting parameter of the plasma frequency $\omega_{pl}$ as a function of pump-probe delay time. (f) Time evolution of the effective mass $m^*$ of electron gas upon intraband THz pumping.

The unpumped near-field MIR response of the NW acquired in the standard nano-FTIR mode at demodulation order $n = 2$ is shown in Figure 1(c). Both the amplitude spectrum $s(\omega) = s_{NW}(\omega)/s_{Si}(\omega)$ and the phase spectrum $\phi(\omega) = \phi_{NW}(\omega) - \phi_{Si}(\omega)$ are normalized to the response of the Si substrate, assuming its permittivity to be constant for the photon energy range of the MIR probe [24]. Due to the interference between the tip-scattered radiation $E_{scat}$ and the reference beam, which is approximately identical to the incident on the tip field $E_{inc}$, the MCT detector measures the homodyne detection intensity $I \propto |E_{scat}||E_{inc}|e^{i(\phi_{scat}-\phi_{inc})}$ rather than just the intensity of the scattered probe $I \propto |E_{scat}|^2$ [25]. Thus, normalizing the NW spectra to the Si reference eliminates the unknown parameters of the amplitude and phase of the incident field $E_{inc}$. In accordance with equation (1) this simplifies modeling the plasmonic response to just two physical parameters. The fitting of the experimental data provides us with the plasma frequency $\omega_{pl0} = (1050 \pm 10) \text{ cm}^{-1}$ and the damping $\gamma_{el0} = (260 \pm 20) \text{ cm}^{-1}$. As seen from Figure 1(c), the model can reliably reproduce the experimental amplitude and phase spectra.

Samples under study are grown in such a way that Si dopants are incorporated into InGaAs as a donor [17]. Due to the heavy n-doping of the sample, the observed plasmonic resonance is attributed to the response of electrons in the conduction band. The InGaAs Γ-valley conduction band is nonparabolic and can be approximately described by the equation $\varepsilon_k(1 + \alpha\varepsilon_k) = \hbar^2 k^2/(2m_\Gamma)$ [26] with the Γ-point effective mass $m_\Gamma$ and the nonparabolicity parameter $\alpha = 1/E_g$, where $E_g$ is the band gap [27]. The nonparabolicity leads to a variation of the effective mass with the electron density $m^*(n)$. By solving the equation $ne^2/[\varepsilon_0 \varepsilon_{\text{optic}} m^*(n)] = \omega_{\text{pl0}}^2$ we obtain the unpumped electron density $n_0 = (8.5 \pm 0.2) \times 10^{18}$ cm$^{-3}$ and the distribution function of electrons with a chemical potential of $\varepsilon_F = 265$ meV at 300 K. The value of the average effective mass $m_0^*$ corresponding to the electon density $n_0$ equates to $0.06 m_e$. Knowing the effective mass $m_0^*$ also allows us to convert the fitted value of damping $\gamma_{\text{el0}}$ to the mobility $\mu_0 = (600 \pm 50)$ cm$^2$V$^{-1}$s$^{-1}$.

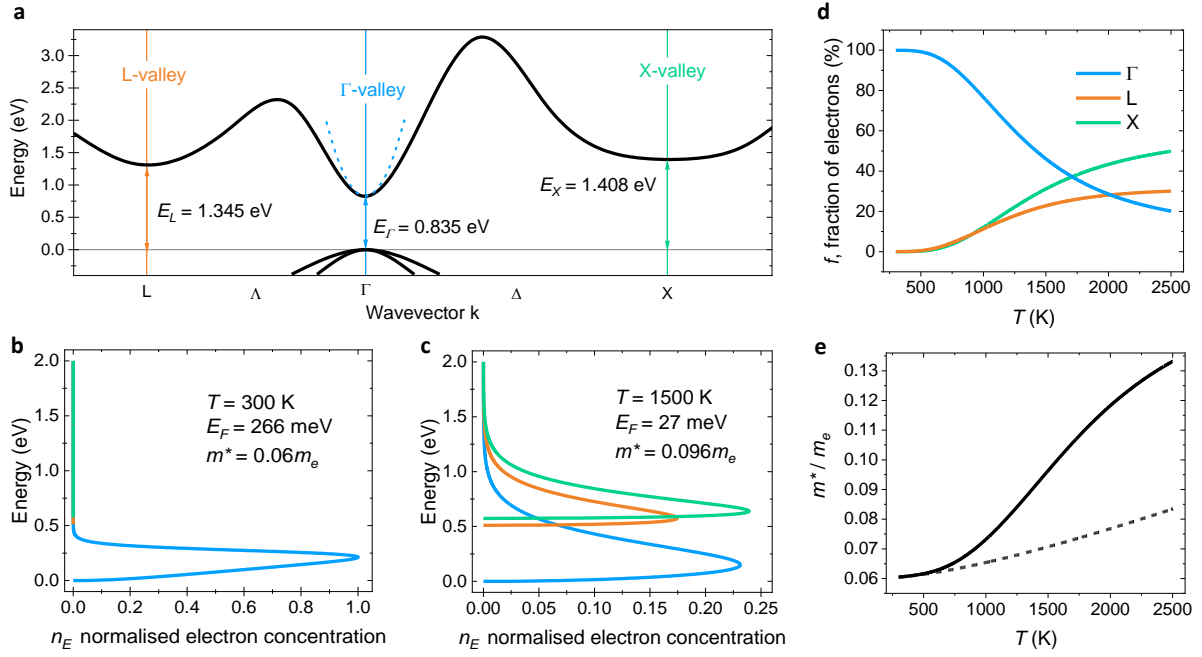

**Fig. 3.** Simulation results: (a) In$_{0.44}$Ga$_{0.56}$As band structure scheme depicting the conduction band valleys relevant to the experiment [34]. Blue dashed line is a parabolic approximation of the nonparabolic Γ-valley. (b, c) Normalized electron distributions for 300 and 1500 K. (d) The calculated fractions of electrons of each conduction band valley versus temperature. (e) Dependence of the total effective mass on temperature. The dashed line represents the simulation neglecting the impact of side valley transfer.

## THz-Pump / MIR-Probe Experiment

For THz-pump / MIR-probe s-SNOM studies, we utilize the intense narrowband THz radiation from FELBE [16] as intraband pump. Its pulses have a repetition rate of 13 MHz and a duration at full width at half maximum (FWHM) of 2.85 ps at the chosen wavelength of λ = 25 µm (12 THz). This is the shortest wavelength that can be efficiently suppressed by a ZnSe high-pass filter placed before the detection scheme to avoid saturation of the detector. The temporal profile of the pulse and the duration were obtained from the measured FEL spectrum using the model in [28]. Unless otherwise mentioned, the data shown in the paper are obtained with an average excitation power of P$_{\text{avg}}$ = 6 mW. The FEL pump beam is guided to the parabolic mirror that focuses it onto the tip apex. The DFG probe source is locked to the sixth harmonic of the FEL repetition rate and synchronized to the FEL pulse train. The time delay between pump and probe pulses is varied by an optical delay line.

The sixfold difference in the repetition rates of the FEL and the DFG source leads to an undesirable situation when the pump pulse excites the sample, and only every sixth probe pulse measures the real response of the sample to this excitation. This can be overcome technically during the measurement process via sideband demodulation technique by introducing an additional modulation of the pump beam to directly detect the photoinduced change in the interferogram [29]. Since the FEL operates in a pulsed mode, our pump is originally modulated at the FEL repetition rate of 13 MHz, which we use as a carrier frequency, while the second harmonic of the AFM cantilever tapping frequency $2\Omega \approx 500$ kHz is utilized as a sideband (see Supporting Information). Nonetheless, the 2 MHz cutoff frequency of the employed MCT detector diminishes the efficiency of the double demodulation technique. Substituting the employed detector with a faster one, however, results in a compromised sensitivity within the probed range. To overcome this issue, we further suggest a data processing approach that enables the extraction of the relevant excited probe pulses from data acquired without additional modulation.

As previously mentioned, when the pump pulse excites the sample, only every sixth probe pulse measures the actual response of the sample to this excitation. The five following probe pulses are idle and measure the response of the sample in an unpumped state until the next pump pulse arrives. Assuming that the detector equally averages intensities of all the received probe pulses, we can define an average-measured intensity $I_{mix}$ as following:

$$I_{mix} = \frac{I_{pumped} + 5 \times I_{unpumped}}{6}, \quad (3)$$

where $I_{pumped}$ and $I_{unpumped}$ are the intensities of the probe pulses scattered from the sample in the excited and unexcited state, respectively. In case of interferometric studies, equation (3) is also valid for each point of the interferogram and, therefore, is applicable to describe the mixing of the pumped and unpumped interference patterns as a whole. Thus, given the interferogram of the sample in the unexcited state we can extract the pumped interferogram from the measured mixed interference pattern $I_{mix}$. Considering the linearity of the Fourier transform, the same procedure can be applied to extract complex pumped scattering spectra from the measured mixed spectra. The extracted pumped spectra exhibit a superior signal-to-noise ratio (SNR) compared to the results obtained via side-demodulation detection (see Supporting Information), and are used for the presentation of the results in the following.

Figures 2(a,b) show near-field amplitude $s(\omega)$ and phase $\phi(\omega)$ spectra of the InGaAs NW obtained with and without THz pumping, extracted according to equation (3) and normalized to the response of Si. Despite the increased noise in the extracted spectra, the implemented fitting model works well and additionally serves as a guide for the eye to observe a red shift of the NW plasma resonance upon the excitation. To trace the full dynamics of the resonance shift we perform a series of spectroscopic scans with a 2 ps step of the optical delay line. Figures 2(c,d) depict the results in a way of a colour map, where every line represents extracted and normalized to Si near-field spectra obtained for different time delays between THz-pump and broadband MIR-probe counted from an arbitrary time zero. One can see a red shift of both amplitude and phase of the

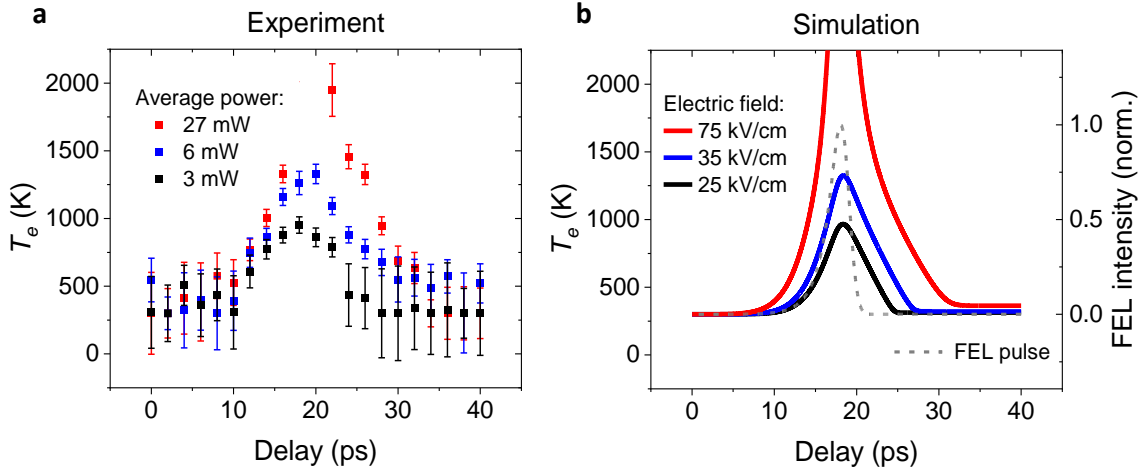

**Fig. 4.** (a) Temporal evolution of the electron-gas temperature upon terahertz-pumping of various powers. (b) Simulation of the temporal evolution of the electron temperature (same axis as part (a)) based on the two-temperature model. Grey dashed line represents normalized intensity profile of the FEL pulse.

near-field plasmonic response and recovery on the same time scale. To quantify the observed shift of the plasma resonance all the spectra are fitted with the point-dipole model and the fitted values of the plasma frequency $\omega_{pl}$ for different time delays are plotted in Figure 2(e). Since the intraband pump does not excite new electrons into the conduction band, the carrier density $n_0 = (8.5 \pm 0.2) \times 10^{18}$ cm$^{-3}$ should remain constant. Thus, the time evolution of the plasma frequency $\omega_{pl}$ can be converted to the time evolution of the effective mass $m^*$ (Figure 2(f)), which quantitatively illustrates the dynamics of the electron gas heating reaching values of up to $0.09 m_e$ for the 6mW THz pump.

## Dynamics of Electron Temperature

Several theoretical [30, 31] and experimental [32, 33] studies are dedicated to the heating of free electrons by mid-infrared radiation. They associate the increase in effective mass of highly doped *n*-GaAs both with the nonparabolicity of the Γ-valley and, at high temperatures, with the transfer of electrons to the heavier mass L-valley, while the small contribution of X-valley electrons is typically neglected. In the case of the In$_{0.44}$Ga$_{0.56}$As alloy, a schematic band diagram of which is depicted in Figure 3(a), the energy separation between the L- and X- minima almost disappears [34]. Therefore, in addition to the nonparabolicity (the parabolic approximation of the Γ-valley is shown in Figure 3(a) as a dashed line) we have to consider the redistribution of heated electrons between the three conduction valleys. By taking into account the boundary condition of carrier density conservation, we calculate electron distributions for various temperatures of the heated electron gas. Figures 3(b,c) depict the calculated electron distributions for 300 and 1500 K, respectively, showing the change of the valley populations and of the Fermi–Dirac distribution.

To quantify the repopulation of hot electrons and its effect on the effective mass we introduce the value of the fraction of electrons in the i-th valley, $f_i$, normalized to the total density of electrons $n_0$. Figure 3(d) shows the calculated temperature dependence of the fraction of electrons in each valley. Initially, all the electrons stay within the Γ-valley, and in order to transfer just 1% of them to the side valleys, the electron gas must be heated above 500 K. As the heating increases, the tail of the distribution widens enough for faster repopulation of electrons, which are evenly distributed over two side valleys up to 1000 K. At higher temperatures, the fraction of electrons settled in X-valley becomes greater than for the L-valley due to the fact that the density-of-states effective mass of X-valley electrons is somewhat larger than that of L-valley electrons [34]. Contribution of electrons of each valley to the total effective mass is estimated as for conductivity problems:

$$\frac{1}{m^*} = \frac{f_X}{m_X^*} + \frac{f_L}{m_L^*} + \frac{f_\Gamma}{m_\Gamma^*(n_\Gamma)}, \qquad (4)$$

where $f_i$ is the fraction of electrons in the corresponding valley, $m_\Gamma^*(n_\Gamma)$ is the effective mass in the Γ-valley considering the nonparabolicity, $m_X^*$ and $m_L^*$ are optical effective masses for anisotropic X- and L-minima characterized by longitudinal $m_l$ and transverse $m_t$ effective masses and defined as $(m)^{-1} = (m_l^{-1} + 2m_t^{-1})/3$. As a result, we obtain a simulated dependence of the total effective mass $m^*(T)$ on temperature shown in Figure 3(e). To emphasize the influence of side-valley electrons, we additionally simulated the temperature dependence of the total effective mass $m^*$ considering that all electrons stay within the nonparabolic Γ-valley [Figure 3(e), dashed line]. At the temperatures below 500 K both curves are equal and the effective mass is rising because of the Γ-valley nonparabolicity, while further heating leads to a significant increase of the total effective mass due to the transfer of electrons to the side valleys.

Knowing the temperature dependence of the total effective mass $m^*(T)$ enables us to extract the time evolution of the electron temperature $T(t)$ from the experimental results on the temporal changes in the effective mass $m^*(t)$. Figure 4(a) represents the dynamics of the electron temperature $T(t)$ upon THz-pumping of various powers. The data points related to the excitation power of 6 mW are extracted directly from the dataset depicted in Figure 2(f) replacing each mass value $m^*$ with the associated temperature $T$ according to the simulated $m^*(T)$ dependence. The data points corresponding to other power excitations are acquired in the same manner and detailed outcomes of the related pump-probe experiments are available in the Supporting Information. In Figure 4(a), all data points within the temperature range below 500 K exhibit significant error bars, attributed to error propagation resulting from the relatively flat segment of the simulated $m^*(T)$ dependence. At higher temperatures the slope of the $m^*(T)$ is larger, which makes it possible to track temperature changes more precisely. However, excessively high temperatures are also challenging to track. Exciting the sample with a power of 27 mW enables us to discern temperatures up to approximately 2000 K as the upper detection limit. Excessive heating causes the plasma resonance to be shifted beyond our MIR probing range and the exact temperature of the electron gas remains unknown.

**Two-Temperature Modelling**

For interpretation of the temporal response shown in Figure 4(a) we performed a modeling of the electron temperature dynamics using the two-temperature model [35]:

$$C_e(T_e)\frac{dT_e}{dt} = P_{\text{FEL}}(t) - g(T_e - T_l), \tag{5}$$

$$C_l\frac{dT_l}{dt} = g(T_e - T_l), \tag{6}$$

where $T_e$ and $T_l$ are the temperatures of the electrons and the lattice, respectively; $C_e$ and $C_l$ are the specific heats of the electron and the lattice subsystems, respectively; $g$ is the electron-phonon coupling constant; and $P_{\text{FEL}}(t)$ is the FEL power density absorbed by the electrons. The latter function can be written as

$$P_{\text{FEL}}(t) = \sigma_1(\omega_{\text{FEL}}, T_e(t))E_{\text{FEL}}^2(t), \tag{7}$$

where $\sigma_1(\omega_{FEL}, T_e) = \omega_{\text{FEL}}\text{Im}[\varepsilon_{\text{NW}}(\omega_{FEL}, T_e)]\varepsilon_0$ is the optical conductivity of the NW at the FEL frequency $\omega_{\text{FEL}}$ estimated from the Drude model parameters in Eq. (2), and $E_{\text{FEL}}$ is the local electric field in the NW under the AFM tip induced by the FEL pumping. Since the effective mass depends on the transient electronic temperature $T_e(t)$ as depicted in Figure 3(e), the plasma frequency $\omega_{\text{pl}}$ and the optical conductivity in the Drude model also vary with time in our simulation. In contrast to metals, where the Fermi energy is large, the condition $k_BT \ll \varepsilon_F$ is not fulfilled in our case of a doped semiconductor. Therefore, the usual assumption that the electronic specific heat $C_e$ is proportional to $T_e$ does not hold. In our modeling we used the $C_e(T_e)$ dependence calculated for the actual band structure of $In_{0.44}Ga_{0.56}As$ (see Supporting Information for details).

Figure 4(b) shows the simulation results for three peak electric-field amplitudes of the FEL radiation that properly scale with the FEL powers used in the experiment. We obtain reasonably good agreement with the experimentally measured temperature dynamics by varying only two parameters: the electron-phonon coupling $g$ and the conversion factor between the FEL power and the local electric field $E_{\text{FEL}}^2$. The simulation confirms the quicker rise and slower decay of the electron temperature and accurately reproduces the substantial power-dependent extension in the duration the system stays in a heated state compared to the FEL pulse duration. The model captures the peak temperature levels for two lower excitation fields, but the simulated temperature for the highest field of 75 kV/cm seems to be overestimated. Although the experimental data also reveal that the electron temperature exceeds 2000 K around the peak for the FEL power of 27 mW, the simulation shows even more extreme heating with maximal $T_e$ of 3700 K.

First, let us discuss the electron-phonon coupling. The coupling coefficient used in the simulation is independent of $T_e$ and equals $g \approx 10^{14}$ J/(m³ K s). It results in temperature relaxation times of 2…5 ps for the electronic temperature between 500 and 2000 K. This agrees with theoretical results predicting a nearly temperature-independent electron cooling time around 1 ps for electronic temperatures up to 500 K [36]. We would like to point out that our assumption of the temperature-independent electron-phonon coupling does not take into account that the LO phonon emission rate for the valley electrons is much higher than for the Γ-valley [37]. This would lead to an increase of the electron-phonon coupling $g$ at high electronic temperatures, consequently accelerating the cooling dynamics. However, this effect should not drastically affect the temperature dynamics in the range of electronic temperatures detected in our experiment.

Second, we compare the estimated peak FEL field in the focus of the parabolic with the simulation results. For the focal length of 11 mm and the measured FEL beam diameter of ≈4 mm the size of the diffraction-limited spot is ≈70 μm. Thus, knowing the FEL pulse parameters and considering the Fresnel transmission coefficient of ≈25% for the p-polarized pump, we can estimate the peak FEL field inside the NW samples to be ≈15 kV/cm for an FEL power of 3 mW. Taking into account uncertainties in the experimental and modeling parameters, this value nearly matches the local electric field $E_{FEL}$ used in the simulations. This fact demonstrates that the pump-induced THz near-field under the SNOM tip does not have superior effect on the sample heating compared to the focused far-field. At the first glance, this result contradicts the near-field modeling predicting a few orders of magnitude enhancement of the near-field at the sample surface [38, 39]. However, the normal electric field component *inside* the sample is reduced by factor |ε| with respect to the field *outside*. In our case, the permittivity of the doped NW at the FEL frequency can be estimated from equation (2) giving $|\varepsilon_{NW}| \approx 60$. Thus, it is possible that the tip enhancement of the near-field is compensated by the plasma screening inside the NW, resulting in the comparable effect on the sample heating from both near- and far-field.

## CONCLUSION

Our investigation of heavily-doped GaAs/InGaAs core-shell NWs via THz-pump / MIR-probe s-SNOM provides valuable insights into the heating and cooling dynamics of electron gas in an individual nanowire. The localized plasmon resonance experiences a strong red-shift under THz-FEL excitation. We have demonstrated that this behavior can be explained by the increase of the averaged effective electron mass caused by the hot electron distribution that populates side valleys of the conduction band in InGaAs. Additionally, the simulation of heated electron distributions has allowed to estimate the temporal evolution of electron-gas temperatures under THz pumping. The observed features of the temporal response were accurately reproduced by the two-temperature model, yielding estimated values for both the electron-phonon coupling and the local pump field. The results provide relaxation times that are consistent with theoretical predictions and suggest that the pump-induced THz near field under the SNOM tip has a minimal impact on sample heating when compared to the focused far-field. Importantly, THz-pump / MIR-probe s-SNOM has proven itself as a powerful tool for quantitatively exploring the nonlinear interaction of nanostructures with THz radiation, further broadening its applicability in the field of nanoscale science and technology.

## ASSOCIATED CONTENT
## Supporting Information

The results of the sideband modulation technique along with the point-dipole model based simulations, the extended power-dependent results, and the calculation of the specific heat of the electron gas (PDF).

# AUTHOR INFORMATION

**Corresponding Authors**

**Andrei Luferau** – *Institute of Ion Beam Physics and Materials Research, Helmholtz-Zentrum Dresden-Rossendorf, Dresden 01328, Germany; Institut für Angewandte Physik, Technische Universität Dresden, Dresden 01187, Germany;*
E-mail: a.luferau@hzdr.de

**Alexej Pashkin** – *Institute of Ion Beam Physics and Materials Research, Helmholtz-Zentrum Dresden-Rossendorf, Dresden 01328, Germany;*
E-mail: a.pashkin@hzdr.de

**Authors**

**Maximilian Obst** – *Institut für Angewandte Physik, Technische Universität Dresden, Dresden 01187, Germany; Würzburg-Dresden Cluster of Excellence - EXC 2147 (ct.qmat), Dresden 01062, Germany;*

**Stephan Winnerl** – *Institute of Ion Beam Physics and Materials Research, Helmholtz-Zentrum Dresden-Rossendorf, Dresden 01328, Germany;*

**Susanne C. Kehr** – *Institut für Angewandte Physik, Technische Universität Dresden, Dresden 01187, Germany; Würzburg-Dresden Cluster of Excellence - EXC 2147 (ct.qmat), Dresden 01062, Germany;*

**Emmanouil Dimakis** – *Institute of Ion Beam Physics and Materials Research, Helmholtz-Zentrum Dresden-Rossendorf, Dresden 01328, Germany;*

**Felix G. Kaps** – *Institut für Angewandte Physik, Technische Universität Dresden, Dresden 01187, Germany; Würzburg-Dresden Cluster of Excellence - EXC 2147 (ct.qmat), Dresden 01062, Germany;*

**Osama Hatem** – *Institut für Angewandte Physik, Technische Universität Dresden, Dresden 01187, Germany; Würzburg-Dresden Cluster of Excellence - EXC 2147 (ct.qmat), Dresden 01062, Germany; Department of Engineering Physics and Mathematics, Faculty of Engineering, Tanta University, Tanta 31511, Egypt;*

**Kalliopi Mavridou** – *Institute of Ion Beam Physics and Materials Research, Helmholtz-Zentrum Dresden-Rossendorf, Dresden 01328, Germany; Institut für Angewandte Physik, Technische Universität Dresden, Dresden 01187, Germany;*

**Lukas M. Eng** – *Institut für Angewandte Physik, Technische Universität Dresden, Dresden 01187, Germany; Würzburg-Dresden Cluster of Excellence - EXC 2147 (ct.qmat), Dresden 01062, Germany; Collaborative Research Center 1415, Technische Universität Dresden, Dresden 01069, Germany;*

**Manfred Helm** – *Institute of Ion Beam Physics and Materials Research, Helmholtz-Zentrum Dresden-Rossendorf, Dresden 01328, Germany; Institut für Angewandte Physik, Technische Universität Dresden, Dresden 01187, Germany;*

**Notes**

The authors declare no competing financial interest.

## ACKNOWLEDGMENTS

The authors are grateful to J Michael Klopf and the ELBE team for the operation of the free-electron laser FELBE and for dedicated support and to Thales de Oliveira and Xiaoxiao Sun for their experimental assistance at the Helmholtz-Zentrum Dresden-Rossendorf.

A.L. expresses gratitude to Markus B. Raschke for insightful discussions on sideband demodulation technique, conveyed through private communications.

M.O., S.C.K, F.G.K., O.H, and L.M.E acknowledge funding by the Bundesministerium für Bildung und Forschung (BMBF, Federal Ministry of Education and Research, Germany) grant numbers 05K19ODA, 05K19ODB, and 05K22ODA as well as the Deutsche Forschungsgemeinschaft (DFG, German Research Foundation) through project CRC1415 (ID: 417590517) and the Würzburg-Dresden Cluster of Excellence "ct.qmat" (EXC 2147, ID: 390858490).

## REFERENCES

1. P. Kun, P. Parkinson, L. Fu, Q. Gao, N. Jiang, Y.-N. Guo, F. Wang, H. J. Joyce, J. L. Boland, H. H. Tan, C. Jagadish, and M. B. Johnston, "Single nanowire photoconductive terahertz detectors," Nano Lett. **15**(1), 206–210 (2015).
2. P. Kun, D. Jevtics, F. Zhang, S. Sterzl, D. A. Damry, M. U. Rothmann, B. Guilhabert, M. J. Strain, H. H. Tan, L. M. Herz, L. Fu, M. D. Dawson, A. Hurtado, C. Jagadish, and M. B. Johnston, "Three-dimensional cross-nanowire networks recover full terahertz state," Science **368**(6490), 510–513 (2020).
3. P. Parkinson, J. Lloyd-Hughes, Q. Gao, H. Hoe Tan, C. Jagadish, M. B. Johnston, and L. M. Herz, "Transient terahertz conductivity of GaAs nanowires," Nano Lett. **7**(7), 2162–2165 (2007).
4. J. M. Stiegler, A. J. Huber, S. L. Diedenhofen, J. Gomez Rivas, R. E. Algra, E. P. A. M. Bakkers, and R. Hillenbrand, "Nanoscale free-carrier profiling of individual semiconductor nanowires by infrared near-field nanoscopy," Nano Lett. **10**(4), 1387–1392 (2010).
5. F. Kuschewski, H.-G. von Ribbeck, J. Döring, S. Winnerl, L. M. Eng, and S. C. Kehr, "Narrow-band near-field nanoscopy in the spectral range from 1.3 to 8.5 THz," Appl. Phys. Lett. **108**, 11 (2016).
6. D. Lang, L. Balaghi, S. Winnerl, H. Schneider, R. Hübner, S. C. Kehr, L. M. Eng, M. Helm, E. Dimakis, and A. Pashkin, "Nonlinear plasmonic response of doped nanowires observed by infrared nanospectroscopy," Nanotechnology **30**(8), 084003 (2018).
7. H. J. Joyce, C. J. Docherty, Q. Gao, H. Hoe Tan, C. Jagadish, J. Lloyd-Hughes, L. M. Herz, and M. B. Johnston, "Electronic properties of GaAs, InAs and InP nanowires studied by terahertz spectroscopy," Nanotechnology **24**(21), 214006 (2013).

8. I. Fotev, L. Balaghi, J. Schmidt, H. Schneider, M. Helm, E. Dimakis, and A. Pashkin, "Electron dynamics in InxGa1-xAs shells around GaAs nanowires probed by terahertz spectroscopy," Nanotechnology **30**(24), 244004 (2019).
9. M. Eisele, T. L. Cocker, M. A. Huber, M. Plankl, L. Viti, D. Ercolani, L. Sorba, M. S. Vitiello, and R. Huber, "Ultrafast multi-terahertz nano-spectroscopy with sub-cycle temporal resolution," Nat. Photonics **8**(11), 841–845 (2014).
10. M. Wagner, A. S. McLeod, S. J. Maddox, Z. Fei, M. Liu, R. D. Averitt, M. M. Fogler, S. R. Bank, F. Keilmann, and D. N. Basov, "Ultrafast dynamics of surface plasmons in InAs by time-resolved infrared nanospectroscopy," Nano Lett. **14**(8), 4529–4534 (2014).
11. A. Pizzuto, E. Castro-Camus, W. Wilson, W. Choi, X. Li, and D. M. Mittleman, "Nonlocal time-resolved terahertz spectroscopy in the near field," ACS Photonics **8**(10), 2904–2911 (2021).
12. V. Pushkarev, H. Němec, V. C. Paingad, J. Maňák, V. Jurka, V. Novák, T. Ostatnický, and P. Kužel, "Charge Transport in Single-Crystalline GaAs Nanobars: Impact of Band Bending Revealed by Terahertz Spectroscopy," Adv. Funct. Mater. **32**(5), 2107403 (2022).
13. M. Egard, S. Johansson, A.-C. Johansson, K.-M. Persson, A. W. Dey, B. M. Borg, C. Thelander, L.-E. Wernersson, and E. Lind, "Vertical InAs nanowire wrap gate transistors with ft> 7 GHz and fmax> 20 GHz," Nano Lett. **10**(3), 809–812 (2010).
14. J.-P. Colinge, C.-W. Lee, A. Afzalian, N. D. Akhavan, R. Yan, I. Ferain, P. Razavi, B. O'Neill, A. Blake, M. White, A.-M. Kelleher, B. McCarthy and R. Murphy, "Nanowire transistors without junctions," Nat. Nanotechnol. **5**(3), 225–229 (2010).
15. S. Morkötter, N. Jeon, D. Rudolph, B. Loitsch, D. Spirkoska, E. Hoffmann, M. Döblinger, S. Matich, J. J. Finley, L. J. Lauhon, G. Abstreiter, and G. Koblmüller, "Demonstration of confined electron gas and steep-slope behavior in delta-doped GaAs-AlGaAs core--shell nanowire transistors," Nano Lett. **15**(5), 3295–3302 (2015).
16. M. Helm, S. Winnerl, A. Pashkin, J. M. Klopf, J.-C. Deinert, S. Kovalev, P. Evtushenko, U. Lehnert, R. Xiang, A. Arnold, A. Wagner, S. M. Schmidt, U. Schramm, T. Cowan, and P. Michel, "The ELBE infrared and THz facility at Helmholtz-Zentrum Dresden-Rossendorf," Eur. Phys. J. Plus **138**(2), 158 (2023).
17. E. Dimakis, M. Ramsteiner, A. Tahraoui, H. Riechert, and L. Geelhaar, "Shell-doping of GaAs nanowires with Si for n-type conductivity," Nano Res. **5**, 796–804 (2012).
18. L. Balaghi, G. Bussone, R. Grifone, R. Hübner, J. Grenzer, M. Ghorbani-Asl, A. V. Krasheninnikov, H. Schneider, M. Helm, and E. Dimakis, "Widely tunable GaAs bandgap via strain engineering in core/shell nanowires with large lattice mismatch," Nat. Commun. **10**(1), 2793 (2019).
19. F. Huth, A. Govyadinov, S. Amarie, W. Nuansing, F. Keilmann, and R. Hillenbrand, "Nano-FTIR absorption spectroscopy of molecular fingerprints at 20 nm spatial resolution," Nano Lett. **12**(8), 3973–3978 (2012).
20. R. Hillenbrand, B. Knoll, and F. Keilmann, "Pure optical contrast in scattering-type scanning near-field microscopy," J. Microsc. **202**(1), 77–83 (2001).
21. F. Huth, M. Schnell, J. Wittborn, N. Ocelic, and R. Hillenbrand, "Infrared-spectroscopic nanoimaging with a thermal source," Nature Mater. **10**(5), 352–356 (2011).
22. B. Knoll and F. Keilmann, "Enhanced dielectric contrast in scattering-type scanning near-field optical microscopy," Opt. Commun. **182**(4-6), 321–328 (2000).
23. A. Cvitkovic, N. Ocelic, and R. Hillenbrand, "Analytical model for quantitative prediction of material contrasts in scattering-type near-field optical microscopy," Opt. Express **15**(14), 8550–8565 (2007).
24. D. Chandler-Horowitz and P. M. Amirtharaj, "High-accuracy, midinfrared ($450 cm^{-1} \leqslant \omega \leqslant 4000 cm^{-1}$) refractive index values of silicon," J. Appl. Phys. **97**(12), (2005).
25. M. Eisele, Ultrafast multi-terahertz nano-spectroscopy with sub-cycle temporal resolution, Doctoral dissertation, University of Regensburg, 2015.
26. M. Lundstrom, *Fundamentals of Carrier Transport* (Cambridge University Press, Cambridge, 2000).
27. B. R. Nag, *Electron Transport in Compound Semiconductors* (Springer-Verlag, Berlin, 1980)
28. S. Regensburger, S. Winnerl, J. M. Klopf, H. Lu, A. C. Gossard, and S. Preu, "Picosecond-scale terahertz pulse characterization with field-effect transistors," IEEE Trans. Terahertz Sci. Technol. **9**(3), 262–271 (2019).

29. J. Nishida, S. C. Johnson, P. T. S. Chang, D. M. Wharton, S. A. Dönges, O. Khatib, and M. B. Raschke, "Ultrafast infrared nano-imaging of far-from-equilibrium carrier and vibrational dynamics," Nat. Commun. **13**(1), 1083 (2022).
30. G. Shkerdin, J. Stiens, and R. Vounckx, "Comparative study of the intra- and intervalley contributions to the free-carrier induced optical nonlinearity in n-GaAs," J. Appl. Phys. **85**(7), 3807–3818 (1999).
31. G. Shkerdin, J. Stiens, and R. Vounckx, "A multi-valley model for hot free-electron nonlinearities at 10.6 μm in highly doped n-GaAs," Eur. Phys. J. Appl. Phys. **12**(3), 169–180 (2000).
32. K. Kash, P. A. Wolff, and W. A. Bonner, "Nonlinear optical studies of picosecond relaxation times of electrons in n-GaAs and n-GaSb," Appl. Phys. Lett. **42**(2), 173–175 (1983).
33. S. Y. Auyang and P. A. Wolff, "Free-carrier-induced third-order optical nonlinearities in semiconductors," JOSA B **6**(4), 595–605 (1989).
34. I. Vurgaftman, J. R. Meyer, and L. R. Ram-Mohan, "Band parameters for III--V compound semiconductors and their alloys," J. Appl. Phys. **89**(11), 5815–5875 (2001).
35. S. I. Anisimov, B. L. Kapeliovich, T. L. Perelman, "Electron emission from metal surfaces exposed to ultrashort laser pulses," Zh. Eksp. Teor. Fiz. **66**(2), 375–377 (1974);
36. S.-C. Lee, I. Galbraith, and C. R. Pidgeon, "Influence of electron temperature and carrier concentration on electron-LO-phonon intersubband scattering in wide GaAs/AlxGa1-xAs quantum wells," Phys. Rev. B **52**(3), 1874 (1995).
37. J.-J. Zhou and M. Bernardi, "Ab initio electron mobility and polar phonon scattering in GaAs," Phys. Rev. B **94**, 201201(R) (2016).
38. F. Mooshammer, M. A. Huber, F. Sandner, M. Plankl, M. Zizlsperger, and R. Huber, "Quantifying nanoscale electromagnetic fields in near-field microscopy by Fourier demodulation analysis," ACS Photonics **7**(2), 344–351 (2020).
39. F. Huth, A. Chuvilin, M. Schnell, I. Amenabar, R. Krutokhvostov, S. Lopatin, and R. Hillenbrand, "Resonant antenna probes for tip-enhanced infrared near-field microscopy," Nano Lett. **13**(3), 1065–1072 (2013).

# Hot electron dynamics in a semiconductor nanowire under intense THz excitation

## Supporting Information

### 1. Sideband modulation technique

The sixfold difference in the operating frequencies of the FEL and the DFG can be overcome by detecting the photoinduced change in the probe signal via the modulation of the pump beam. Since the FEL operates in a pulsed mode, our pump is originally modulated at the FEL repetition rate ($\Omega_M$ = 13 MHz). Considering the probe signal is initially demodulated at the second harmonic ($2\Omega \approx 500$ kHz) of the AFM cantilever's tapping frequency, we use the FEL repetition rate $\Omega_M$ = 13 MHz as the carrier frequency and detect the the pump-induced change of the probe signal as a sideband at $\Omega_M \pm 2\Omega$.

To perform sideband modulated THz-pump / MIR-probe s-SNOM studies, we additionally incorporate a Zurich Instruments lock-in into the measurement scheme, which processes the signal from the MCT detector at the sideband frequency. Supplementary Figure 1(a) represents the THz-pump induced interferograms of the GaAs/InGaAs core-shell NW for various time delays between THz-pump and broadband MIR-probe, counted from an arbitrary time zero corresponding to the expected arrival of the pump pulse. While pump and probe pulse doesn't overlap, there is no any pump-induced signal being detected. At delay of 2 ps we can see the signal starting to emerge from the background noise. Further overlap of pump and probe pulses leads to increase of the amplitude of the fringes, its saturation, decay and vanishing.

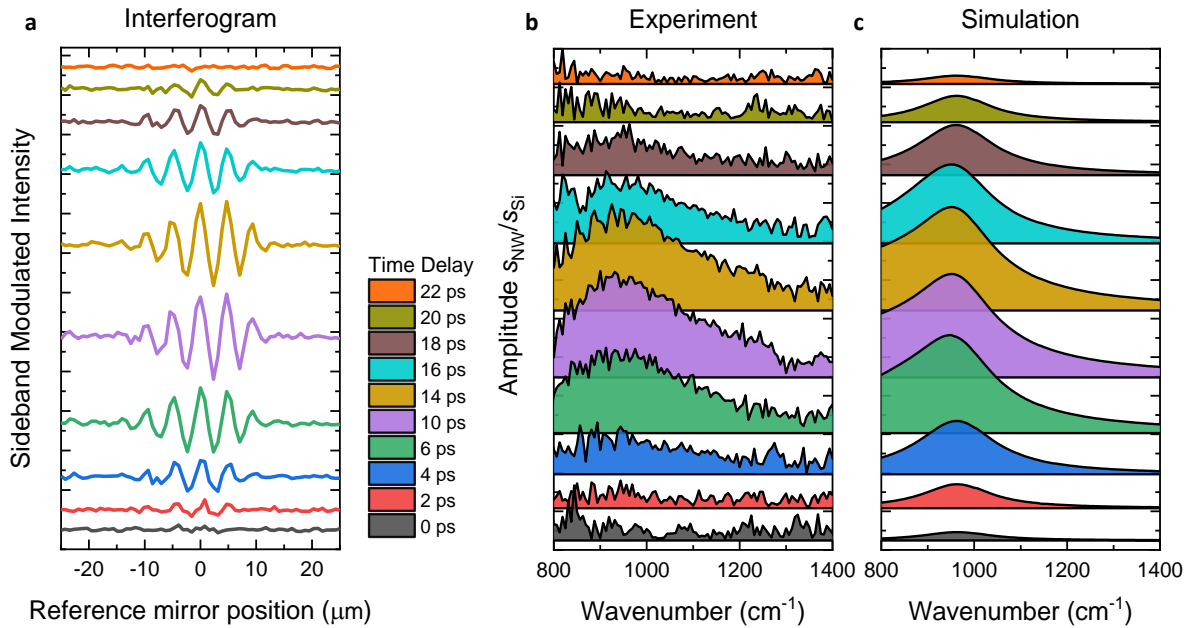

**Supplementary Fig. 1.** (a) Sideband-modulated pump-probe interferograms of the GaAs/InGaAs core-shell NW. (b) Spectral evolution of the pump-induced differential amplitude of the plasma resonance in the NW, normalized to the response of a Si. (c) Simulation of pump-induced differential amplitude of the plasma resonance based on the point-dipole model.

The Fourier transform of the interferograms yields the spectral evolution of the pump-induced differential amplitude of the plasma resonance in the NW. Similar to standard nano-FTIR studies, the sideband-modulated spectra need to be normalized to the reference spectra obtained without side modulation. This normalization is crucial to eliminate unknown parameters of the amplitude and phase of the incident field $E_{inc}$ (more details can be found in the paper). Supplementary Figure 1(b) represents the series of the Fourier transformed sideband modulated spectra obtained at different time delays and normalized to the response of a Si substrate.

Incorporating a modification of the point-dipole model discussed in the paper, we simulate the evolution of the differential sideband-modulated spectra as the difference between two plasmonic resonances: the unpumped resonance characterized by defined initial values of the plasma frequency $\omega_{pl0}$ and the damping $\gamma_{el0}$, and the pumped resonance with parameters modified by the pump ($\omega_{pl}(t)$ and $\gamma_{el}(t)$). In our simulation, we treat $\omega_{pl}(t)$ and $\gamma_{el}(t)$ as adjustable parameters for each time delay. As seen from Supplementary Figure 1(c), this approach allows us to reproduce the general trends in our data. While the simulation of the low-amplitude spectra are essentially an approximation owing to the difficulty of discerning the signal from the background noise, the high-amplitude saturated spectra also represent a form of approximation. The issue arises from the pump-induced shift of the plasma resonance to the edge or beyond our MIR probing range, resulting in the saturation of the peak in the differential spectra. Further shifts of the pumped resonance primarily result in changes to the shape of the left shoulder of the peak, which we cannot properly trace because of the heightened noise levels at the edge of our MIR probe range.

It's worth to mention that we also tried to perform the sideband modulation via mechanical chopping of the pump beam with a frequency of several kHz and detecting the sideband a $2\Omega \pm \Omega_M$. This approach provided us with an unsatisfactory noise level, which could be caused by optical disturbances of the tip at a frequency below the tapping frequency $\Omega$.

Additionally, it is important to highlight that the observed pump-induced shift of the plasma resonance to the edge of the probing range is consistent with our power-dependent results shown in the following section. However, the present average excitation power ($P_{avg}$ = 10 mW) cannot be directly compared, as it constitutes a distinct measurement. Parameters such as beam path alignment, pulse shape, and tip sharpness may differ in this context.

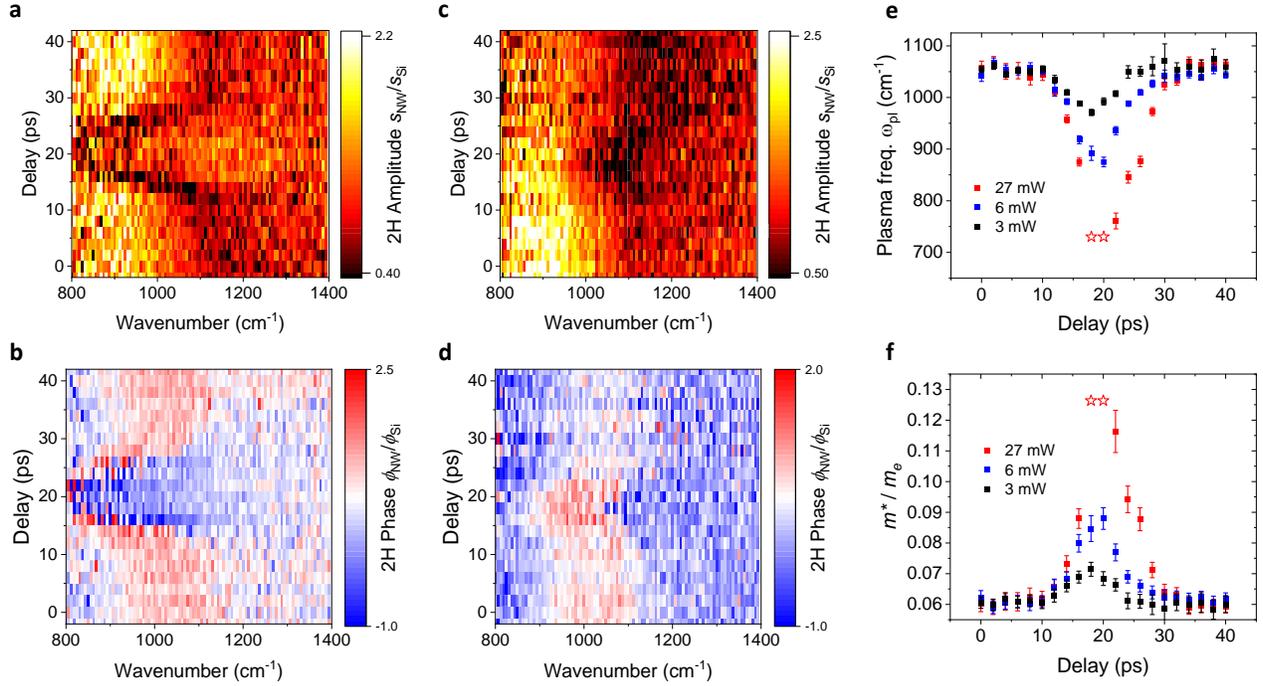

**Supplementary Fig. 2.** (a, b, c, d) Color maps illustrating the evolution of the near-field amplitude s(ω) and the phase ϕ(ω) spectra of doped InGaAs NW upon THz photoexcitation at $P_{avg}$ = 27 mW (a, b) and $P_{avg}$ = 3 mW (c, d), normalized to the response of Si substrate. Every line represents near-field spectra obtained for different time delays between THz-pump and broadband mid-infrared probe. (e) Fitting parameter of the plasma frequency $\omega_{pl}$ as a function of pump-probe delay time. (f) Time evolution of the effective mass $m^*$ of electron gas upon intraband THz pumping at various power levels.

## 2. Power-dependent results

Supplementary Figures 2(a,b) and 2(c,d) represents the color maps illustrating the evolution of the near-field amplitude s(ω) and the phase ϕ(ω) spectra of doped the GaAs/InGaAs core-shell NW upon THz photoexcitation at $P_{avg}$ = 3 mW and $P_{avg}$ = 27 mW respectively. Every line depicts near-field spectra extracted in accordance with equation (3) and obtained at various time delays between the THz-pump and MIR probe. All the power-dependent data, including the results presented in the paper, were obtained under identical conditions, differing only in the intensity of the pump beam, and subsequently normalized to the same reference response of Si.

Exciting the sample with a power of 3 mW causes a faintly discernible shift of the plasma resonance, while 27mW excitation results in the plasma resonance shifting beyond our MIR probing range. To quantify the observed shift of the plasma resonance all the spectra are fitted with the point-dipole model and the fitted values of the plasma frequency $\omega_{pl}$ for different time delays are plotted in Supplementary Figure 2(e) along with data from Figure 2(e) of the paper. The data points corresponding to the plasma resonance being shifted beyond the probing range are highlighted with star symbols. Given that the carrier density remains constant during intraband pumping, we convert the time evolution of the plasma frequency $\omega_{pl}$ into the time evolution of the effective mass $m^*$ (Supplementary Figure 2(f)).

## 3. Specific heat of the electron gas in the $In_{0.44}Ga_{0.56}As$ shell

The conduction band of $In_{0.44}Ga_{0.56}As$ is highly non-parabolic. In addition to the non-parabolicity of the Γ-valley, the side valleys strongly affect the total electron energy at elevated temperatures. Using the band structure shown in Fig. 3a we calculated the total electron energy for temperatures between 300 and 5000 K (Supplementary Fig. 3(a)). The specific heat is calculated as the derivative of the electron energy with respect to temperature. The resulting curve is shown in Supplementary Fig. 3(b). One can see that the initial (nearly) linear increase of the specific heat with temperature break down around 1000 K. The specific heat reaches its maximum around 1150 K and starts to decrease at higher temperatures as a result of the increasing electron population in the L- and X-valleys.

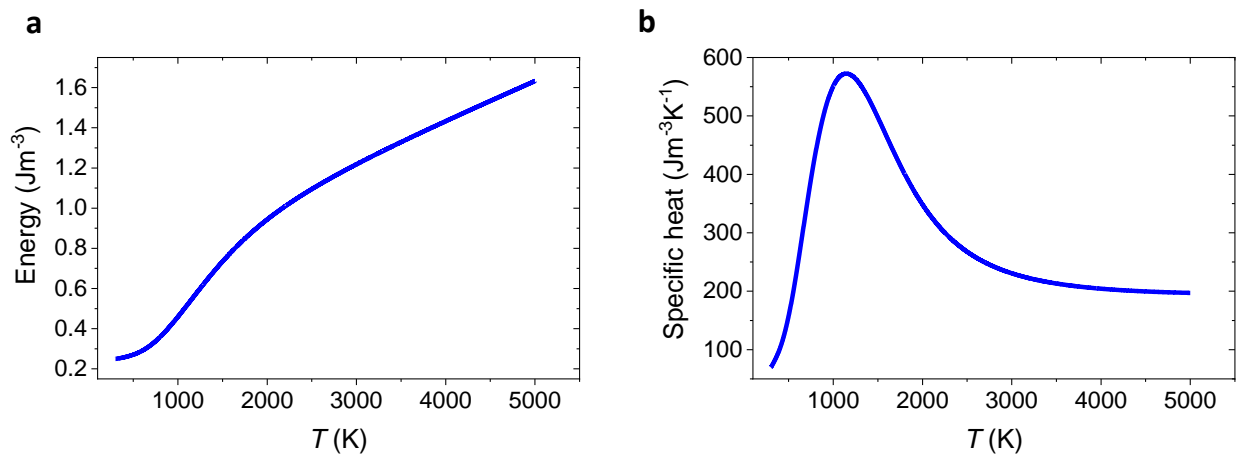

**Supplementary Fig. 3.** (a) Total electron energy in the conduction band of InGaAs as a function of temperature. (b) Electronic specific heat as a function of temperature used in the two-temperature modeling.